\documentclass[
reprint,
superscriptaddress,
aps,
prb,
]{revtex4-2}
\usepackage{comment}	
\usepackage{graphicx}	
\usepackage{dcolumn}	
\usepackage{bm}			
\usepackage{hyperref}	
\usepackage{amsmath}	
\usepackage{amssymb}	
\usepackage{tikz,xcolor,hyperref}

\usepackage{xcolor}
\definecolor{lightgraybg}{gray}{0.85}  

\newcommand{\orcid}[1]{\hspace{0.8mm}\href{https://orcid.org/#1}{\raisebox{-0.05\height}{\includegraphics[height=0.7em]{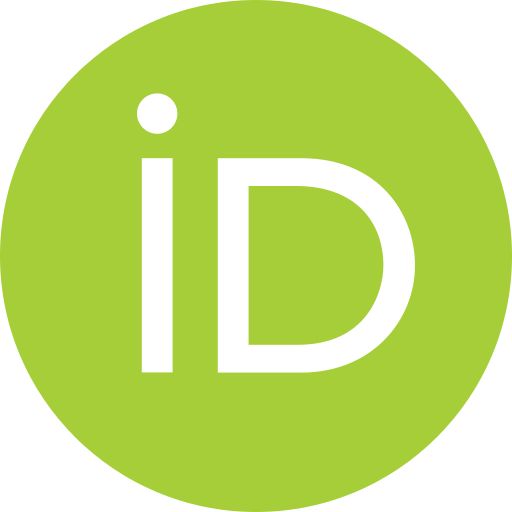}}}}

\newcommand{\new}[1]{\textcolor{black} {#1}}
\begin{document}
\title{Pair anisotropy in disordered magnetic systems}
\author{K. Das\orcid{0000-0002-8137-4148}}
\affiliation{\href{https://ror.org/000sfad56}{Institute of Physics}, Polish Academy of Sciences, Aleja Lotnikow 32/46,\\ PL-02668, Warsaw, Poland.}
\author{N. Gonzalez Szwacki\orcid{0000-0002-0518-844X}}%
\affiliation{Faculty  of Physics, \href{https://ror.org/039bjqg32}{University of Warsaw}, Ludwika Pasteura 5,\\ PL-02093, Warsaw, Poland.}
\author{K. Gas\orcid{0000-0003-3722-764X}}
\affiliation{\href{https://ror.org/000sfad56}{Institute of Physics}, Polish Academy of Sciences, Aleja Lotnikow 32/46,\\ PL-02668, Warsaw, Poland.}
\affiliation{Center for Science and Innovation in Spintronics, \href{https://ror.org/01dq60k83}{Tohoku University},\\ 2-1-1 Katahira, Aoba-ku, Sendai 980-8577, Japan.}
\affiliation{Laboratory for Nanoelectronics and Spintronics, Research Institute of Electrical\\ Communication, \href{https://ror.org/01dq60k83}{Tohoku University}, 2-1-1 Katahira, Aoba-ku,\\ Sendai 980-8577, Japan.}
\author{M. Sawicki\orcid{0000-0001-5740-3677}}
\affiliation{\href{https://ror.org/000sfad56}{Institute of Physics}, Polish Academy of Sciences, Aleja Lotnikow 32/46,\\ PL-02668, Warsaw, Poland.}
%
%
\author{R. Hayn\orcid{0000-0002-7659-0073}}%
\affiliation{\href{https://ror.org/0238zyh04}{IM2NP}, UMR 7334, \href{https://ror.org/035xkbk20}{Universite d’Aix-Marseille} and CNRS, Campus\\ St. Jérôme, Case 142, FR-13397, Marseille, France.}
\author{D. Sztenkiel\orcid{0000-0002-4759-7575}}  \email{sztenkiel@ifpan.edu.pl}
\affiliation{\href{https://ror.org/000sfad56}{Institute of Physics}, Polish Academy of Sciences, Aleja Lotnikow 32/46,\\ PL-02668, Warsaw, Poland.}

\begin{abstract}
Accurate modelling of magnetism is pivotal for elucidating the microscopic origins of magnetic phenomena in functional materials. However, for a specified class of materials, such as random dilute ferromagnets or alloys, the reliance on simplifying assumptions, such as single-ion anisotropy, limits the accuracy of existing spin models. In such systems, there is a significant probability of the formation of nearest-neighbor magnetic ion pairs or higher order clusters, whose presence breaks the local symmetry of otherwise isolated magnetic species. Here, we introduce the concept of pair-induced uniaxial anisotropy and demonstrate how nearby atoms influence each other’s anisotropic behavior. This effect is investigated in the dilute magnetic semiconductor Ga$_{1-x}$Mn$_x$N, by means of  density functional theory calculations. The inclusion of pair anisotropy in the atomistic spin simulations significantly improves the agreement between simulated and experimental magnetization curves, in contrast to models that consider only single-ion anisotropy.
\end{abstract}
%
\maketitle
%
%
\section{Introduction}
%
Understanding and controlling magnetic anisotropy is central to the design of functional magnetic materials for spintronic and information storage applications.
In many materials, the magnetic anisotropy is typically modeled using single-ion terms derived under the assumption of isolated magnetic centers in high-symmetry environments. This approach, often based on mapping total energies from density functional theory (DFT) as a function of spin orientation onto an effective spin Hamiltonian, has been successfully applied to a wide range of materials: from transition metal dopants in semiconductors~\cite{Lusakowski:2015_JPhysCM}, metals~\cite{Steiauf:2005_PRB}, to rare-earth systems~\cite{Lee:2025_CompMater}, and low-dimensional magnets~\cite{TellezMora:2024_CompMater, Xu:2018_CompMater}. 
However, in disordered systems such as dilute magnetic semiconductors (DMSs) or random alloys, the single-ion approximation often proves insufficient. The presence of nearest neighbour (NN) magnetic ion pairs and small clusters breaks the local symmetry of otherwise isolated magnetic species and can give rise to anisotropy contributions not captured by single-ion models. 
\new{Specifically, this pair-induced anisotropy stems from two distinct physical mechanisms: the distortion of the local crystalline environment—which modifies the crystal field and the resulting single-ion-like response—and the symmetric part of the anisotropic exchange interaction. While the former reflects how the presence of a second magnetic ion perturbs the local site symmetry, the latter represents an intrinsic directional dependence of the spin-spin coupling itself.}
Our findings suggest that pair-induced anisotropy is a generic feature of magnetically disordered systems, and may prove essential for accurate modeling of spin dynamics in a wide range of materials: from dilute magnetic semiconductors and oxides, 2D van der Waals magnets with randomly distributed spin centers, to spin glasses, and cluster-based magnets.

\new{The investigation of pairwise interactions has long been recognized as a key element in understanding the magnetism of dilute magnetic semiconductors. Our work aligns with and extends this established research direction. A notable example of the significance of pairwise interactions is the study of (Ga,Mn)As \cite{Birowska:2012_PRL}, where the non-random distribution of Mn–Mn pairs was shown to break the cubic symmetry of the host and induce bulk uniaxial anisotropy. Our results for Ga$_{1-x}$Mn$_x$N further extend the importance of these effects. We demonstrate that even in systems where the pairs are randomly distributed and do not alter the global wurtzite symmetry, their contribution to the magnetic anisotropy energy is crucial for a quantitatively accurate description of experimental magnetization data.
Additionally, recent theoretical work further validates the approach of using Mn–Mn pairs as the fundamental units for describing exchange interactions in III-V magnetic semiconductors \cite{Kokurin:2024_PRB}. Our findings expand this perspective by demonstrating that the same pairwise configurations are critical for understanding the system's magnetic anisotropy.}

Here we investigate the microscopic origin of such pair-induced anisotropy in wurtzite Ga$_{1-x}$Mn$_x$N, a prototypical dilute magnetic semiconductor that exhibits a rare combination of insulating and ferromagnetic behavior. In this material, a small fraction of Ga cations in the GaN host lattice is substituted by Mn ions. 
What makes this compound really worthwhile is that the magnetic anisotropy of individual Mn ions can be modulated via the inverse piezoelectric effect, enabling electric-field control~\cite{Sztenkiel:2016_NatComm, Sztenkiel:2025_CommMater}.
This capability opens pathways for the realization of precessional magnetization switching driven by sub-nanosecond electric pulses. Achieving such control mandates a detailed understanding of the magnetic anisotropy mechanisms in this system.
Notably, the Pt/(Ga,Mn)N interface has recently demonstrated spintronic characteristics on par with leading platforms such as YIG/Pt heterostructures, supported by comparable spin mixing conductance values~\cite{MendozaRodarte2024}.

In wurtzite Ga$_{1-x}$Mn$_x$N, each Mn ion has $N=12$ nearest neighbour Ga cation sites, any of which may host another Mn ion. 
\new{To calculate the probability that a given Mn ion has exactly $k$ nearest neighbor Mn ions, we use the binomial distribution $P(k)=\binom{N}{k} x^k (1-x)^{N-k}$. The probability $p$ of a Mn ion participating in any NN Mn cluster is $1-P(0)=1-(1 - x)^{12}$, yielding $p \cong 52\%$ for $x = 6\%$. In particular, the probability that a given Mn ion has exactly one Mn nearest-neighbor is $P(1)=36.5\%$, while the probability that it belongs to a larger cluster is $P(k \geq 2) \cong 16\%$.}
Thus, the single-ion approximation breaks down even at moderate Mn concentrations, underscoring the importance of considering pair induced magnetic anisotropy in systems generally regarded as dilute. Specifically, the presence of a second Mn ion in NN position disrupts the local crystal field symmetry of the first Mn ion, introducing an additional uniaxial anisotropy aligned with the Mn--Mn axis, an effect we refer to as \emph{pair anisotropy}. To substantiate the existence of such anisotropy, we perform DFT simulations on supercells containing two Mn ions in NN configurations. By incorporating spin--orbit coupling and computing total energies for various spin orientations, we extract magnetic anisotropy energies and directions by fitting the results to a simplified spin Hamiltonian.
Finally, we show that including these pair-induced terms in the atomistic spin simulations significantly enhances the agreement between simulated and experimental magnetization curves for representative Ga$_{1-x}$Mn$_x$N sample.

%
%
\section{Results and Discussion}
%
%
\subsection{First-principles results - structural distortions}
%

%
Our first-principles calculations are carried out within the DFT framework, using the projector augmented-wave (PAW) method~\cite{Bloechl1994} and the generalized gradient approximation (GGA) functional as implemented in the Vienna Ab initio Simulation Package (VASP)\cite{Hafner:1997_SP}. The pseudopotentials are sourced from the standard VASP library. For details, please see appendix~\ref{app:dft}.
\begin{figure*}
	\centering
	\includegraphics[width=\linewidth]{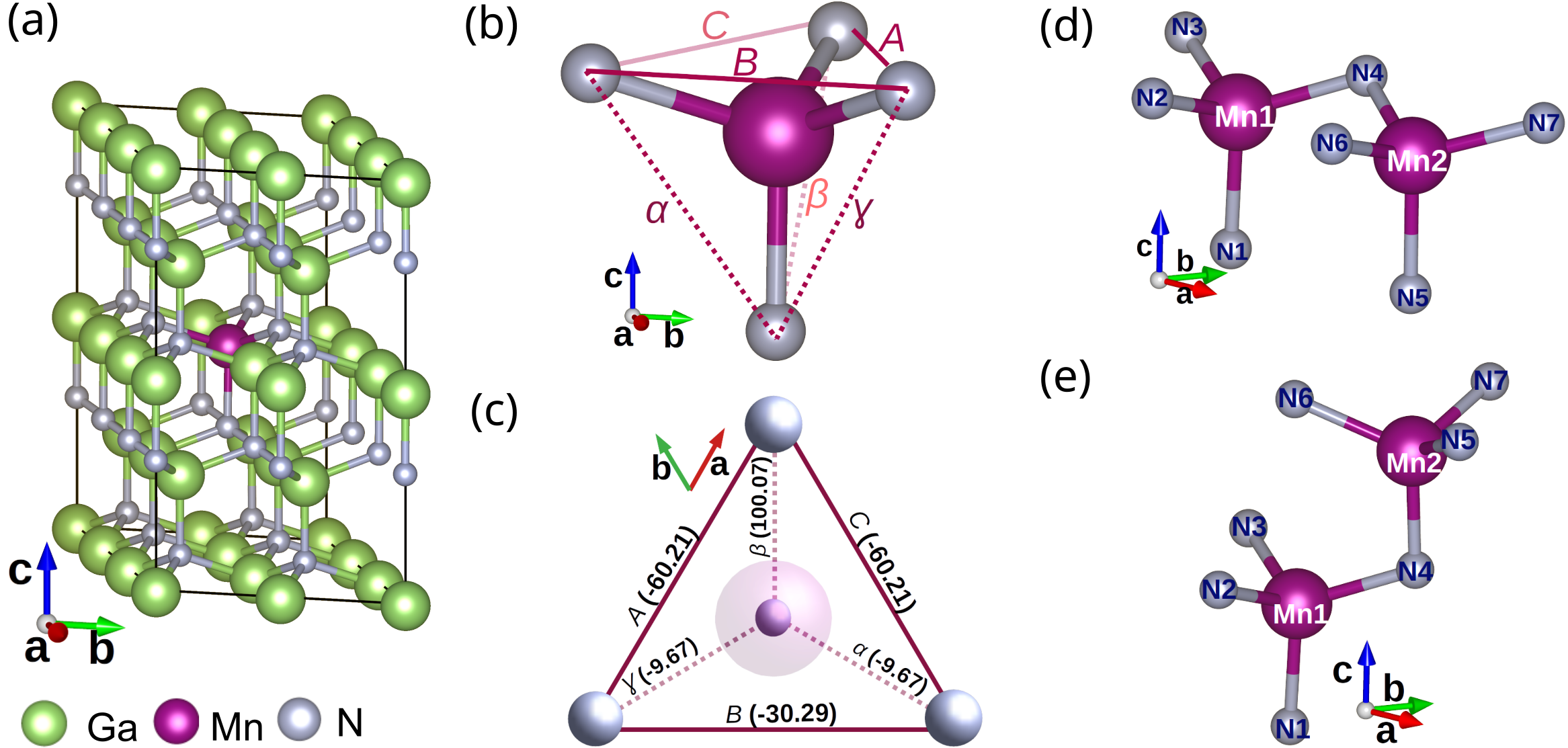}
	\caption{Investigated wurtzite Ga$_{1-x}$Mn$_x$N structure. \textbf{(a)} The $3\times2\times2$ supercell of GaN containing a single Mn ion. \textbf{(b)} Close-up of Mn and its four nearest-neighbor nitrogen anions. Distances labeled $A$, $B$, $C$, $\alpha$, $\beta$, and $\gamma$ represent different N--N separations. \textbf{(c)} Top view of (b). Parenthetical values indicate changes in these distances due to trigonal and Jahn-Teller distortions in pm (see Eq.~\ref{eq:jt_para} in the main text). \textbf{(d)} and \textbf{(e)} A single Mn ion is surrounded by 12 nearest-neighbor Ga cations, which can be potential sites for other Mn substitution. This geometry gives rise to six symmetry-equivalent Mn--Mn in-plane pairs (example in \textbf{d}) and six out-of-plane pairs (example in \textbf{e}). Panels (d) and (e) illustrate Mn ions and their nearest-neighbor nitrogen environments.}
	\label{fig:allsupercell}
\end{figure*}

\begin{figure*}
	\centering
	\includegraphics[width=1.0\linewidth]{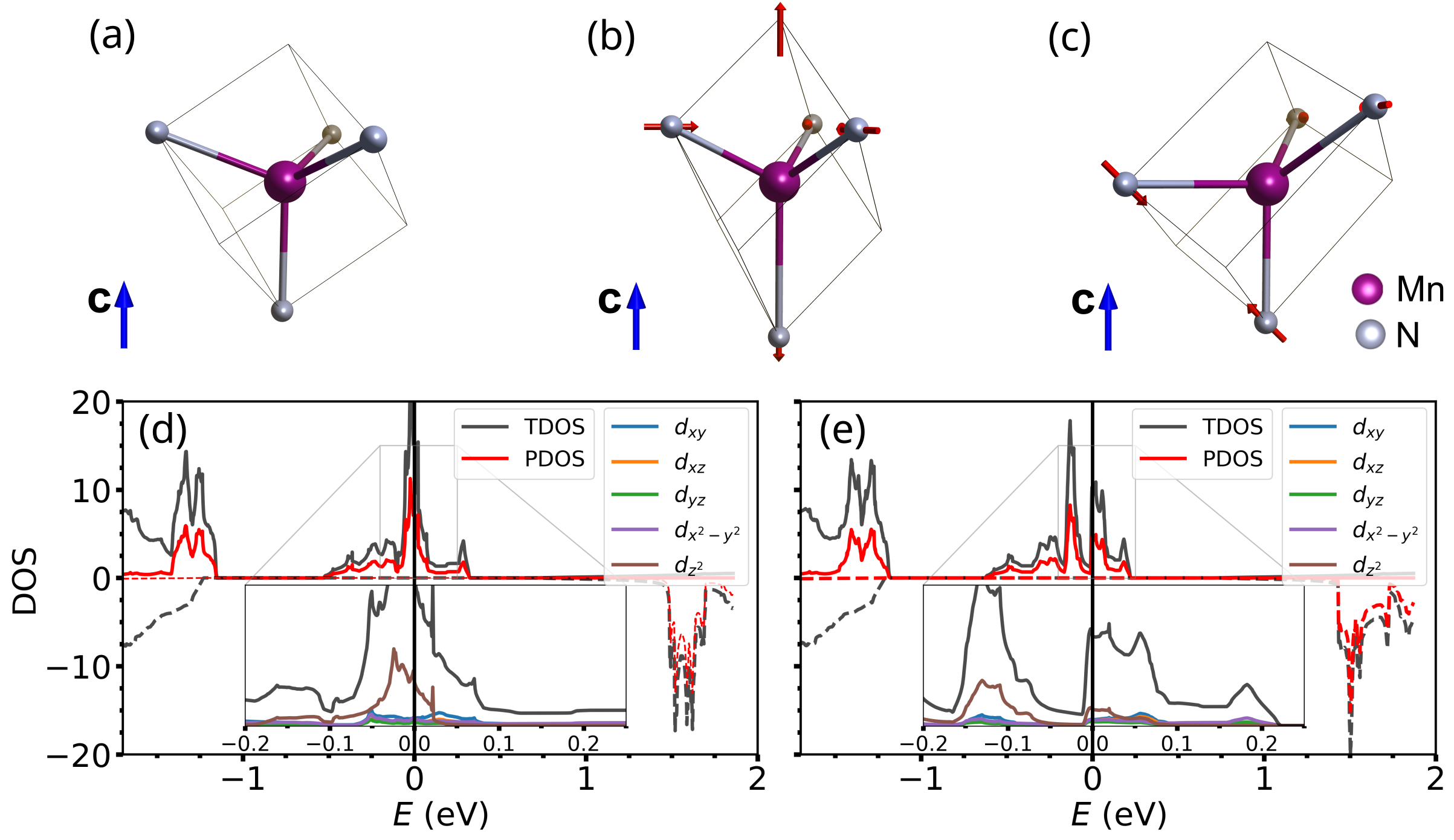}
	\caption{Trigonal and Jahn--Teller (JT) distortions and their effect on the density of state (DOS) calculations. (\textbf{a}) Ideal tetrahedral configuration of the Mn ion with its four nearest-neighbor nitrogen anions (no trigonal or JT distortion). Dashed lines are guides for the eye in illustrating the original cubic symmetry, which is modified in panels (b) and (c). (\textbf{b}) Trigonal distortion along the \textbf{c}-axis of GaN. (\textbf{c}) Jahn-Teller distortion reducing the symmetry from tetrahedral to tetragonal. (\textbf{d}) Partial density of states (PDOS) of the Mn ion and total density of states (TDOS) of the supercell with one Mn ion in wurtzite GaN with optimized GaN atomic positions and lattice parameters (only trigonal distortion present). (\textbf{e}) PDOS and TDOS after full relaxation of both atomic positions and lattice parameters, incorporating Jahn-Teller distortion. Colored lines in the inset of (d, e) show PDOS contributions from different Mn $3d$ orbitals. Solid and dashed lines represent spin-up and spin-down states, respectively. The black vertical line at $E = 0$ indicates the Fermi level.\label{fig:mn_dos}}	
\end{figure*}

As a starting point, we investigate supercell with one Mn ion in wurtzite GaN, shown in Fig.~\ref{fig:allsupercell}~\textbf{a}.
We confirm that the main anisotropy terms associated with the Mn$^{3+}$ ion ($d^4$ configuration) consist of the intrinsic trigonal anisotropy imposed by the wurtzite crystal structure, as well as the Jahn--Teller (JT) effect. A schematic representation of these deformations is provided in Figs.~\ref{fig:mn_dos} \textbf{a}-\textbf{c}.
Their quantitative assessment involves analyzing the Mn-centered N--N distances.  
The trigonal distortion aligned with the $c$-axis of the wurtzite lattice either increases or decreases (depending on the sign of distortion) the N--N lengths labeled by $A$, $B$, and $C$, while simultaneously altering $\alpha$, $\beta$, and $\gamma$ in the opposite manner. N--N distance labels are provided in Fig.~\ref{fig:allsupercell}~\textbf{b} and listed in Table~\ref{tab:mono_mn_abc}.
According to the presented data, the GaN supercell exhibits only a trigonal distortion, where $A = B = C$ and $\alpha = \beta = \gamma$.
In contrast, the single Mn-ion configuration reveals the coexistence of both trigonal and JT distortions. The latter one, lowers the local symmetry from tetrahedral to tetragonal, as shown in Fig.~\ref{fig:mn_dos}~\textbf{c}. At any given moment, the JT distortion favors one of three possible cubic directions. Starting from an ideal tetrahedral configuration where all N--N distances are equal, the JT distortion results in the elongation of two N--N distances while compressing the remaining four, or vice versa. In a chosen case, the distance $B$ is larger than both $A$ and $C$, and similarly, $\beta$ exceeds $\alpha$ and $\gamma$, indicating a $D_{2d}$-type distortion around the Mn ion. These distortions can be parameterized using the following expressions:
\begin{equation}
\begin{aligned}
A = C &= d + t - j_1, \\
B     &= d + t + j_1, \\
\alpha = \gamma &= d - t - j_2, \\
\beta  &= d - t + j_2,
\end{aligned}
\label{eq:jt_para}
\end{equation}
where $d \cong 3.252$~\AA~represents the baseline N--N distance in an ideal tetrahedral geometry. The trigonal distortion is characterized by $t$, while $j_1$ and $j_2$ quantify the JT-induced modifications.
The obtained values clearly indicate that the distortions in Ga$_{1-x}$Mn$_x$N differ substantially from those in undoped GaN. While GaN exhibits only a mild trigonal distortion ( $t=0.0073$~\AA, $j_1=j_2=0$ ), the single-Mn-doped system experiences both trigonal and JT distortions ( $t=-0.0453$~\AA, $j_1=0.0150$~\AA, and $j_2=0.0548$~\AA ).
\begin{table}[htb]
	\caption{\label{tab:mono_mn_abc} Distances between nitrogen atoms surrounding the Ga ion for GaN supercell and surrounding the single-Mn ion in the GaN supercell. N--N distance labels are provided in Fig.~\ref{fig:allsupercell}~\textbf{b}}
	\begin{ruledtabular}
		\begin{tabular}{lcccccc}
			N--N distance & $A$ & $B$ & $C$ & $\alpha$ & $\beta$ & $\gamma$ \\
			Unit & (\AA) & (\AA) & (\AA) & (\AA) & (\AA) & (\AA) \\
			\colrule
			undoped GaN  & 3.2174 & 3.2174 & 3.2174 & 3.2100 & 3.2100 & 3.2100 \\
			Mn in GaN    & 3.2273 & 3.2572 & 3.2252 & 3.2100 & 3.3197 & 3.2057 \\
		\end{tabular}
	\end{ruledtabular}
\end{table}

The density of states (DOS) results provide profound insight into how various structural distortions affect the electronic properties of investigated material. Fig.~\ref{fig:mn_dos}~\textbf{d} shows the DOS for a supercell with one Mn ion calculated using the relaxed lattice parameters of Ga$_{1-x}$Mn$_x$N ($|\mathbf{a}|$=$|\mathbf{b}|$=3.217~\AA~and $|\mathbf{c}|$=5.246~\AA) i.e., including only the trigonal distortion. The absence of an energy gap at the Fermi level results in a semi-metallic character.
In Fig.~\ref{fig:mn_dos}~\textbf{e}, the system is allowed to relax atomic positions, thereby introducing a JT distortion.
This additional relaxation results in an energetically favorable configuration with an energy lowering of $\Delta E = -35.3$~meV, consistent with the results of Virot \textit{et al.}~\cite{Virot2010}. 
The static tetragonal JT distortion reduces the local symmetry and splits the highest energy peak.
However, the system does not converge to an insulating ground state, at odds with experimental observations. This shortcoming is typical for standard DFT within the GGA or the local density approximation, which underestimates electronic correlations and therefore fails to reproduce the insulating character expected for Mn$^{3+}$ ions in a wurtzite GaN ~\cite{Kronik2002,Sanyal2003,Sandratskii2004}. While methods such as DFT+U or hybrid functionals can improve the description of electronic localization and open the gap \cite{Stroppa2009, Virot2010,Djermouni:2020_TEPJB}, here we deliberately employ the Perdew--Burke--Ernzerhof (PBE) approximation. The reason is that our study focuses on magnetocrystalline anisotropy, requiring noncollinear spin calculations with spin--orbit coupling and systematic rotations of the Mn spin moments. For such energy evaluations, PBE remains the most robust and widely implemented approach, whereas DFT+U introduces additional complexities and inconsistencies in noncollinear spin--orbit frameworks. Our approach thus prioritizes consistency and comparability of total-energy differences over the absolute accuracy of the electronic gap.
%
%
%
%
%

Next, we investigate systems containing two Mn ions, both in the in-plane and out-of-plane configurations.
First, we introduce a second Mn atom at the NN cation site along the crystallographic $\mathbf{a}$ lattice vector to create an isolated in-plane Mn--Mn pair, as illustrated in Fig.~\ref{fig:allsupercell}~\textbf{d}. To avoid the formation of Mn chains due to periodic boundary conditions, we use a $3 \times 2 \times 2$ supercell. 
To simulate both single and paired Mn ions within the same Ga$_{1-x}$Mn$_x$N layer, we preserve the lattice parameters from the mono-doped supercell and relax only the internal atomic positions. This partial relaxation results in a total energy reduction of $\Delta E = -88.0$~meV.
%
%
%
\begin{table*}[t]
	\caption{\label{tab:mn_n_distance}%
		Nearest-neighbor distances between nitrogen anions (N) and the first (Mn1) and second (Mn2) Mn ions in the in-plane and out-of-plane Mn--Mn pair supercell. Atom labels are defined in Figs.~\ref{fig:allsupercell}\,(d,e). All values are in \AA.}
	\begin{ruledtabular}
		\begin{tabular}{lccrlccr}
			\multicolumn{4}{c}{In-plane pair} & \multicolumn{4}{c}{Out-of-plane pair} \\ 
			\cline{1-4}\cline{5-8}
			Mn1--N1 & Mn1--N2 & Mn1--N3 & Mn1--N4 & Mn1--N1 & Mn1--N2 & Mn1--N3 & Mn1--N4 \\
			2.0037 & 2.0167 & 1.9707 & 1.9355  & 1.9705 & 2.0236 & 2.0096 & 1.9400 \\
			Mn2--N5 & Mn2--N6 & Mn2--N7 & Mn2--N4 & Mn2--N5 & Mn2--N6 & Mn2--N7 & Mn2--N4 \\
			2.0107 & 2.0094 & 1.9676 & 1.9304 & 1.9980 & 1.9752 & 2.0119 & 1.9164 \\
		\end{tabular}
	\end{ruledtabular}
\end{table*}
The evaluation of the structural distortions introduced by the second Mn ion does not reveal any presence of JT distortion. 
\new{The corresponding $N$-$N$ distance tables for Mn pairs are included in Appendix~\ref{app:no_JT} to illustrate the quenching of the Jahn-Teller symmetry.}
We attribute this to the local symmetry breaking caused by the presence of the second Mn ion, which likely induces additional lattice distortion and effectively quenches the JT effect.
To quantify this symmetry breaking, we focus on the Mn--N distances for both Mn atoms, particularly the bond involving the shared nitrogen atom labeled N4 in Fig.~\ref{fig:allsupercell}~\textbf{d}. The distances, provided in Table~\ref{tab:mn_n_distance}, reveal that the Mn--N4 bond is significantly shorter compared to the other Mn--N distances. This asymmetry is expected to generate additional magnetic anisotropy in the system, thereby invalidating the single-ion magnetic anisotropy model in this configuration.

\new{We emphasize that the Jahn-Teller distortion can only occur if the ground state of the system is orbitally degenerate. The absence of JT effect in the case of Mn–Mn pairs is not caused by a lack of driving force for structural relaxation, but rather by the fundamental change in local symmetry.  While the trigonal crystal field of the wurtzite host allows for a degenerate ground state in isolated Mn ions \cite{Gosk2005, Stefanowicz:2010_PRB, Sztenkiel:2020_NJP, Sztenkiel:2022_JMMM}, the proximity of the second Mn ion lifts the orbital degeneracy of the Mn $d$-states, which is a necessary condition for the JT effect to occur. Consequently, the structural relaxations we observe in the Mn-Mn system, such as the Mn–N bond shortening, are not manifestations of the JT effect but are instead symmetry-dictated distortions aligned with the pair axis.}

Finally, we construct an out-of-plane configuration by substituting the top NN Ga atom with Mn ion, as illustrated in Fig.~\ref{fig:allsupercell}~\textbf{e}.
The $3 \times 2 \times 2$ wurtzite supercell provides sufficient spatial separation to ensure that the Mn--Mn pair remains isolated in this geometry.
As in the in-plane case, the lattice parameters are fixed to those of the relaxed mono-doped configuration, and only the ionic positions are allowed to relax. This ionic relaxation leads to a total energy reduction of $\Delta E = -98.2$~meV. In this configuration, we also observe that the Mn--N4 bond is significantly shorter compared to the other Mn--N distances, as shown in Table~\ref{tab:mn_n_distance}. The shortening of the Mn--N4--Mn bonds is expected to introduce an additional uniaxial anisotropy along the Mn--Mn direction. 

\begin{figure*}
	\centering
	\includegraphics[width=14cm, height=6.0cm]{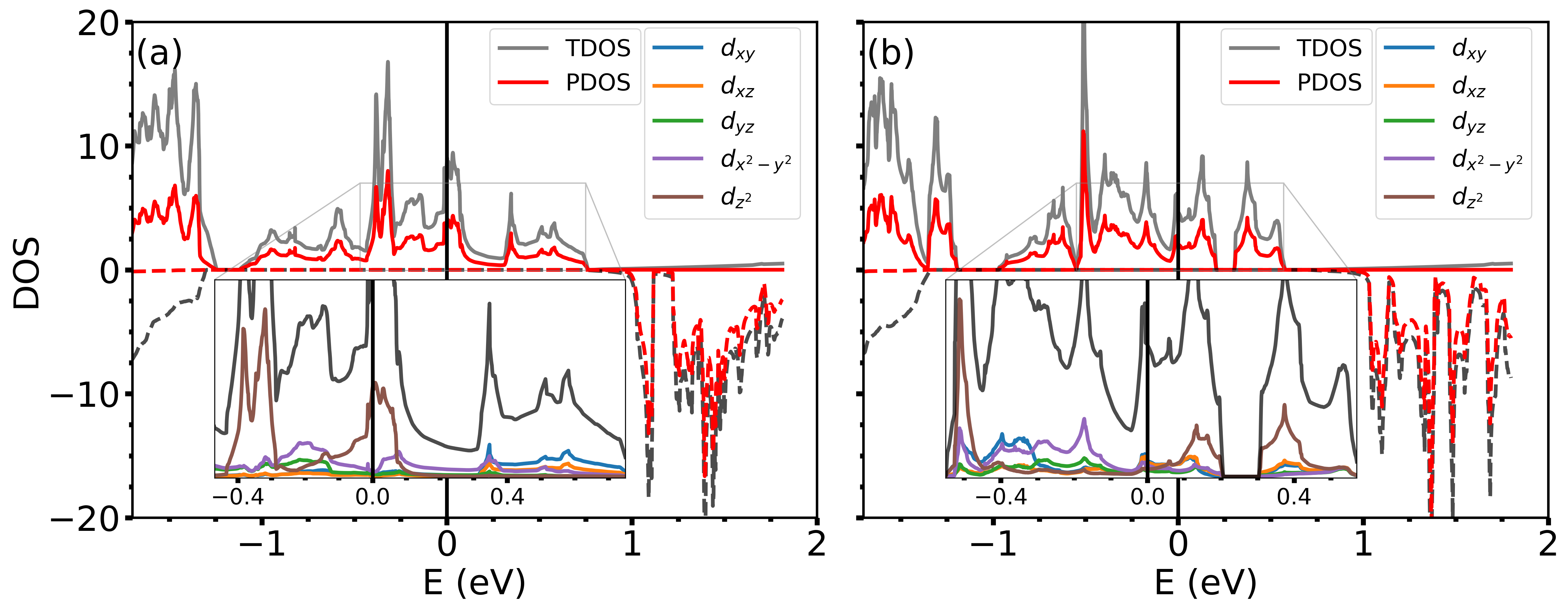}
	\caption{Partial density of states (PDOS) and total density of states (TDOS) for Mn–Mn
		pairs in Ga$_{1-x}$Mn$_x$N. (a) In-plane Mn–Mn pair. (b) Out-of-plane Mn–Mn pair. Solid lines represent spin-up states, dashed lines represent spin-down states. The black vertical line at E = 0 denotes the Fermi level. The colored insets show the PDOS contributions from the different Mn $3d$ orbitals.\label{fig:pair_dos}}
\end{figure*}

The density of states (DOS) calculations presented in Fig.~\ref{fig:pair_dos}~\textbf{a} and \ref{fig:pair_dos}~\textbf{b} reveal a complex electronic structure for both the in-plane and out-of-plane Mn pair configurations. No well-defined energy gap is observed at the Fermi level, indicating that the system resides at the boundary between insulating and metallic behavior. 
%
%
%
\subsection{First-principles results - magnetic anisotropy}
%
%
%
%
%
\begin{figure}[htb]
	\centering
	\includegraphics[width=0.96\linewidth]{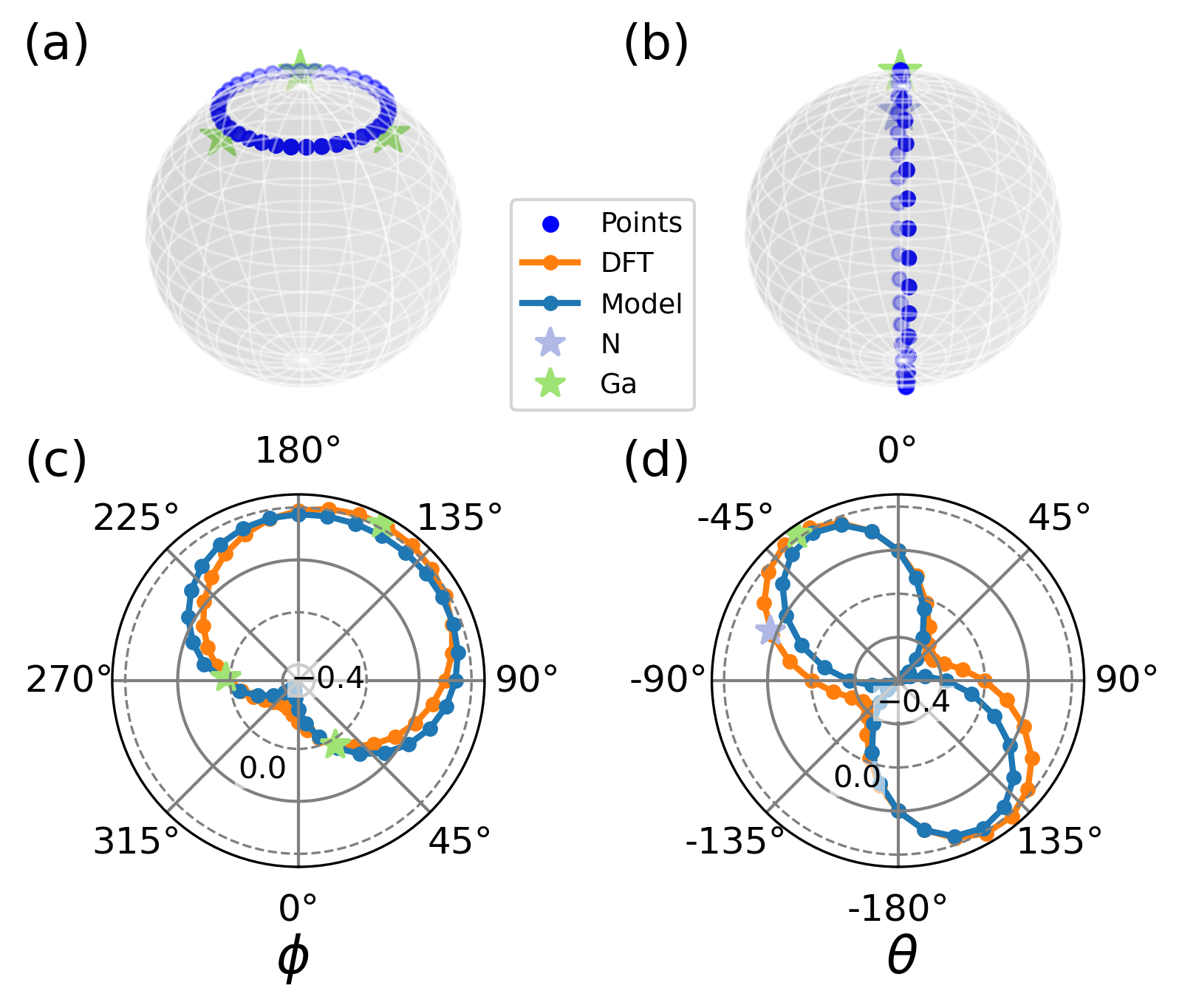}
	\caption{Magnetocrystalline energy of the supercell containing a single Mn ion. Panels (\textbf{c}) and (\textbf{d}) show the energy variation $\Delta E(\theta,\phi)$ obtained from DFT (orange lines) as a function of spherical angles $\phi$ and $\theta$ of Mn spin, respectively. Blue lines represent fits to the spin Hamiltonian model defined in Eq.~\ref{eq:ani_model}. Mn spin orientations are illustrated in (\textbf{a}) and (\textbf{b}), corresponding to rotations over $\phi$ and $\theta$, respectively.\label{fig:mono_fit}}
	
\end{figure}

To investigate the influence of structural distortions on the magnetocrystalline anisotropy, we employ noncollinear spin DFT calculations using the PAW method~\cite{Hobbs2000}, as implemented in VASP. The magnetic moments of Mn ions are rotated systematically through spherical angles $\theta$ and $\phi$, with the $z$-axis of our coordinate system chosen parallel to the $c$-axis of the GaN crystal ($\mathbf{c} \parallel \mathbf{z}$), and the total energy $E(\theta,\phi)$ is computed. For Mn--Mn pairs, both spins are rotated in parallel, effectively suppressing energy contributions from exchange coupling. The anisotropy energy is defined as $\Delta E(\theta, \phi) = E(\theta,\phi) - E_c$, where $E_c$ is the energy for spin alignment along the crystallographic $\textbf{c}$-axis.
To interpret the results, we compare the calculated energy landscape with a phenomenological spin Hamiltonian. For a single Mn ion, the Hamiltonian includes contributions from both trigonal and JT distortions:
\begin{equation}
\label{eq:ani_model}
\mathcal{H} = - \frac{1}{2} K^{TR} S_z^2 - K^{\mathrm{JT}} (\mathbf{S}\cdot\mathbf{e}^{\mathrm{JT}})^2 
\end{equation}
where $K^{TR}$ and $K^{\mathrm{JT}}$ denote the trigonal and JT anisotropy constants, respectively, and $\mathbf{e}^{\mathrm{JT}}$ is a unit vector along the local JT distortion axis. This axis corresponds to one of three equivalent cubic directions~\cite{Gosk2005,Edathumkandy:2022_JMMM} inclined at $\theta_{\mathrm{JT}}$ = $54.73^{\circ}$.

In the Mn-pair systems, the presence of a second Mn ion disrupts the local symmetry around the first ion, suppressing the JT distortion. To model this, we introduce an additional uniaxial anisotropy term aligned with the Mn--Mn bond direction $\mathbf{e}^P$, which differs for in-plane and out-of-plane configurations (see Figs.~\ref{fig:allsupercell} \textbf{d,e}). The spin Hamiltonian becomes:
\begin{equation}
\label{eq:ani_model_pairs}
\mathcal{H} = - \frac{1}{2} K^{TR} \sum_{i=1}^2{ S_{i,z}^2} - K^{P} \sum_{i=1}^2{ (\mathbf{S}_i\cdot\mathbf{e}^P)^2}  
\end{equation}
where $K^{P}$ is the pair-induced anisotropy constant.

\begin{figure*}[htb]
	\centering
	\includegraphics[width=\linewidth]{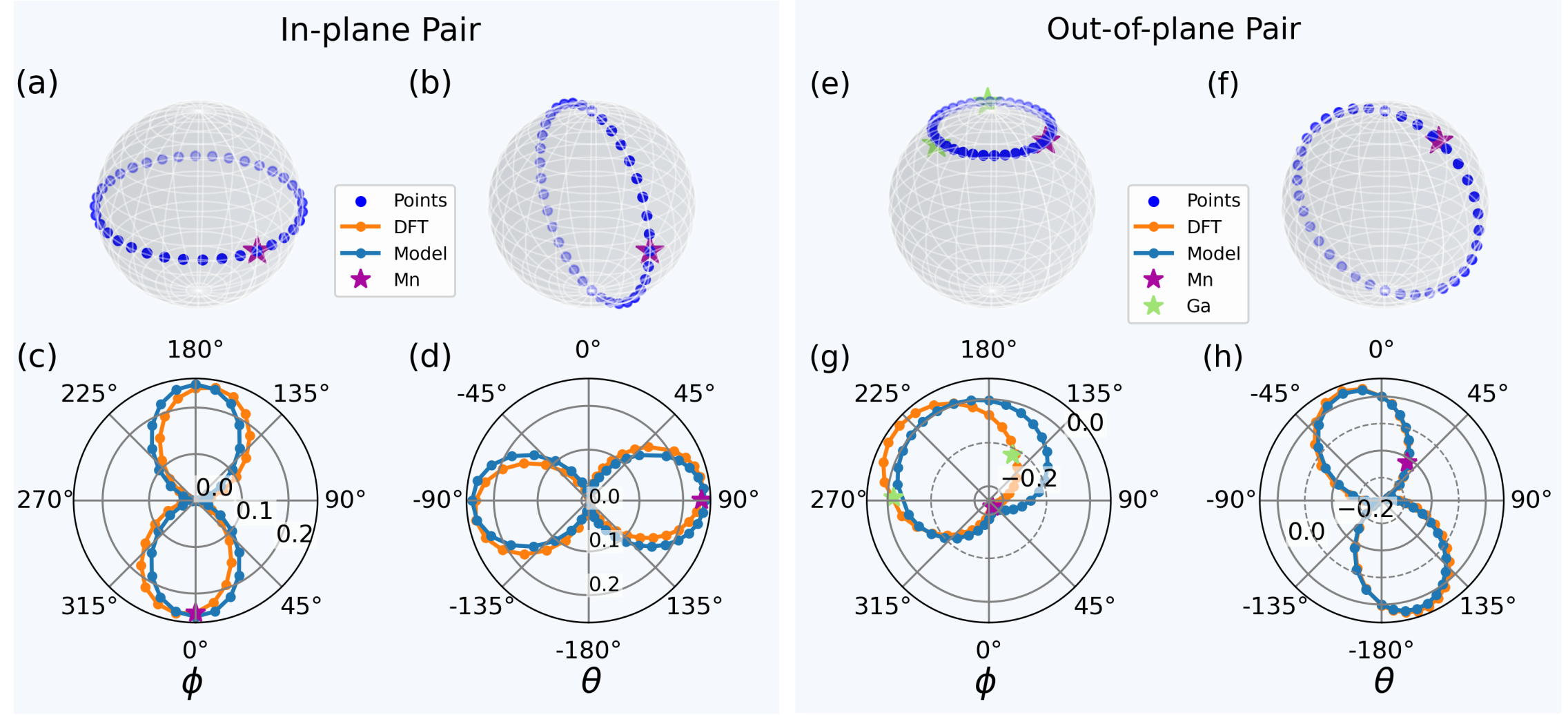}
	\caption{Magnetocrystalline energy of the supercell containing the in-plane (left panel) and out-of-plane (right panel) Mn--Mn pair. DFT results (orange) and spin Hamiltonian fit (blue) for both Mn spins rotations over $\phi$ and $\theta$ are shown in (\textbf{c}-\textbf{d}) and (\textbf{g}-\textbf{h}). The spin Hamiltonian model is defined in Eq.~\ref{eq:ani_model_pairs}. Panels (\textbf{a}), (\textbf{b}), (\textbf{e}) and (\textbf{f}) illustrate the corresponding Mn spins rotation geometries. The position of second Mn ion is represented as a purple star relative to the first one (centered, not shown).\label{fig:pair_fit}}
\end{figure*}
First, we analyze the single-Mn ion supercell. Fig.~\ref{fig:mono_fit}~\textbf{c} shows $\Delta E(\phi)$ with fixed $\theta = 34.65^{\circ}$, corresponding to the spin rotation in a plane containing three Ga nearest neighbors. The energy minimum occurs at $\phi$ = $330^{\circ}$. Varying $\theta$ at fixed $\phi$ = $330^{\circ}$ (Fig.~\ref{fig:mono_fit}~\textbf{d}), we fit the model and extract anisotropy parameters $K^{TR} = -0.3$~meV/ion and $K^{\mathrm{JT}} = 0.68$~meV/ion (with $\mathbf{e}^{\mathrm{JT}}$ along $\theta_{\mathrm{JT}}$ = $54.73^{\circ}$ and $\phi_{\mathrm{JT}}$ = $330^{\circ}$). A very similar value of the Jahn--Teller parameter, $K^{\mathrm{JT}}$=0.75~meV, was reported in Ref. \cite{Edathumkandy:2022_JMMM}, where its magnitude was extracted from crystal-field model simulations.
The fitting to the DFT data in Fig.~\ref{fig:mono_fit} is performed under the conventional assumption that the three Jahn--Teller distortions in the spin Hamiltonian are aligned along the cubic axes, as is standard in previous works \cite{Herbich:1998, Gosk2005, Bonanni:2011_PRB, Sztenkiel:2016_NatComm, Edathumkandy:2022_JMMM, Sztenkiel:2020_NJP, Sztenkiel:2022_JMMM}. This assumption is likewise adopted in all subsequent magnetic simulations. However, an improved agreement is obtained when the Jahn--Teller distortions are instead oriented at an angle of $\theta_{\mathrm{JT}} = 31.46^{\circ}$, rather than strictly cubic (see appendix~\ref{app:non_cube} for details). This result points to a departure from the conventional cubic picture of Jahn-Teller anisotropy, suggesting that local distortions may follow lower-symmetry orientations. If verified experimentally, such a finding would not only refine future magnetic simulations but also reshape our understanding of the Jahn--Teller effect in dilute magnetic semiconductors.

Next, for the in-plane Mn pair (Fig.~\ref{fig:pair_fit}~\textbf{a}-\textbf{d}), the spins are rotated in the plane defined by the Mn-Mn bond vector. Fitting the data yields $K^{TR} = 0.0$~meV/ion and $K^{P} = -0.12$~meV/ion, indicating that the Mn--Mn bond acts as a magnetic hard axis.

For the out-of-plane configuration (Fig.~\ref{fig:pair_fit}~\textbf{e}-\textbf{h}), we obtain $K^{TR} = -0.42$~meV/ion and $K^{P} = 0.12$~meV/ion. The signs indicate that the $c$-axis assumes a hard axis character due to trigonal effects, while the Mn--Mn bond is an easy one. The obtained data clearly show that the presence of a second Mn ion in NN position, introduces an additional uniaxial anisotropy aligned with the Mn--Mn axis. 

Importantly, we have obtained distinct values (signs) of the trigonal anisotropy constant $K^{TR}$ and the pair anisotropy constant $K^{P}$ depending on whether the Mn ions are isolated or form pairs in different geometrical configurations. 
In our opinion, these contrasting results signal that the local structural and electronic environment fundamentally alters the anisotropy. 
Hence, the single-ion anisotropy cannot be naively applied across diverse local arrangements: effective spin models must account for anisotropy parameters that depend on pair geometry. Our results emphasize that Mn--Mn correlations may be essential for capturing the correct magnetic energy landscape in disordered magnetic system with short-range interactions.

%
\subsection{Magnetic anisotropy - discussion}
%

\new{To verify the robustness of our results against electron correlation effects, we performed benchmark calculations of the magnetic anisotropy using the DFT+$U$ approach for $U=2$ and 4~eV. These calculations were conducted for several selected spin directions, using the relaxed atomic positions obtained from the PBE functional. We found that the resulting magnetic anisotropy values are practically identical to those obtained with $U=0$. The details of this comparison are provided in Appendix~\ref{app:Anizo_U}. These results demonstrate that our PBE-based approach is stable and provides a physically sound description of the magnetic anisotropy in Ga$_{1-x}$Mn$_x$N.}

\new{To accurately interpret the origin of the magnetic anisotropy in Mn--Mn pairs, one must distinguish between two physical contributions. The first is the magnetic anisotropy arising from the local crystalline environment, which includes the crystal field of the host lattice and its symmetry breaking induced by the presence of the second Mn ion in the pair. This effect determines the single-ion-like behavior of each Mn spin within the distorted environment. The second contribution is the anisotropic part of the exchange interaction, where the exchange coupling strength itself depends on the orientation of the spin pair relative to the Mn--Mn bond axis. Our DFT results represent the total pair anisotropy, which is the sum of these local environment-induced effects and the intrinsic anisotropic exchange, providing a comprehensive picture of the magnetic energy surface of the Mn-pair. In general, the interaction between two spins $\textbf{S}_1 \cdot \hat{J} \cdot \textbf{S}_2$ is described by a $3 \times 3$  exchange tensor $\hat{J}$  which can be decomposed into three physically distinct terms: the isotropic Heisenberg exchange $ J \textbf{S}_1 \cdot \textbf{S}_2$, the Dzyaloshinskii-Moriya (DM) interaction $\textbf{D}(\textbf{S}_1 \times \textbf{S}_2)$, and the symmetric anisotropic exchange $\Gamma_x S_{1x}S_{2x} + \Gamma_y S_{1y}S_{2y}  + \Gamma_z S_{1z}S_{2z}$  (in the principal axes system, with $\Gamma_x + \Gamma_y + \Gamma_z = 0$) \cite{Shekhtman:1992_PRL}. Our calculations of the pair anisotropy (PA) account for the total change in the system's energy, thus including both the local crystal-field effects and the symmetric anisotropic components of $\Gamma$.}

\new{Importantly, in our DFT procedure both Mn spins are rotated collinearly and in parallel. Under such a constraint, the DM term vanishes identically because $\textbf{S}_i \times \textbf{S}_j=0$. However, our results indicate that the pair anisotropy is dominated by the symmetric anisotropic exchange. Specifically, the calculated energy difference $(\Gamma_x - \Gamma_y)S^2 = 0.204$~meV accounts for the majority of the total pair anisotropy 0.25~meV (see Fig.~\ref{fig:pair_fit}), leaving only a minor contribution to the local crystal-field effects. The calculated exchange parameters are (in meV): $J=75$, $\Gamma_x$=0.028, $\Gamma_y$=-0.023, $\Gamma_z$=-0.0045 and $D_x$= -0.11, $D_y$=-0.025, $D_z$=0.074. While it is well-known that standard DFT functionals tend to overestimate the absolute values of the exchange integrals in nitrides \cite{Sawicki:2012_PRB}, the qualitative picture remains consistent. The fact that the anisotropic part of the exchange tensor is the primary driver of the PA suggests that the magnetic axis orientation is intrinsically linked to the Mn--Mn coupling geometry rather than just the single-ion response of the individual impurities.}

\new{It should be noted that the epitaxial Ga$_{1-x}$Mn$_x$N thin film investigated in this work is grown on a GaN buffer layer, which constrains the in-plane lattice parameters ($a$ and $b$) to those of the buffer. In our DFT calculations, we have adopted these constrained values to accurately reproduce the experimental conditions (a detailed discussion is provided in Appendices A and B). In this context, one must distinguish between global and local anisotropy contributions: while the epitaxial strain or even the Mn concentration $x$ primarily affect the global symmetry and modify the uniaxial (trigonal) anisotropy of the system \cite{Sztenkiel:2016_NatComm, Sztenkiel:2025_CommMater}, the pair anisotropy investigated here is a local effect. This PA arises from the symmetry breaking induced by the Mn--Mn pair geometry. Consequently, the strain may lead to minor quantitative shifts in the PA values, but the fundamental nature of the pair-induced effects remain qualitatively robust and intrinsic to the local ion configuration.}
%
%
\subsection{Magnetic Simulations}\label{Sec:MagneticAnisotropy}
%
%
%
%
\begin{figure}[h]
	\centering
	\includegraphics[width=0.95\linewidth]{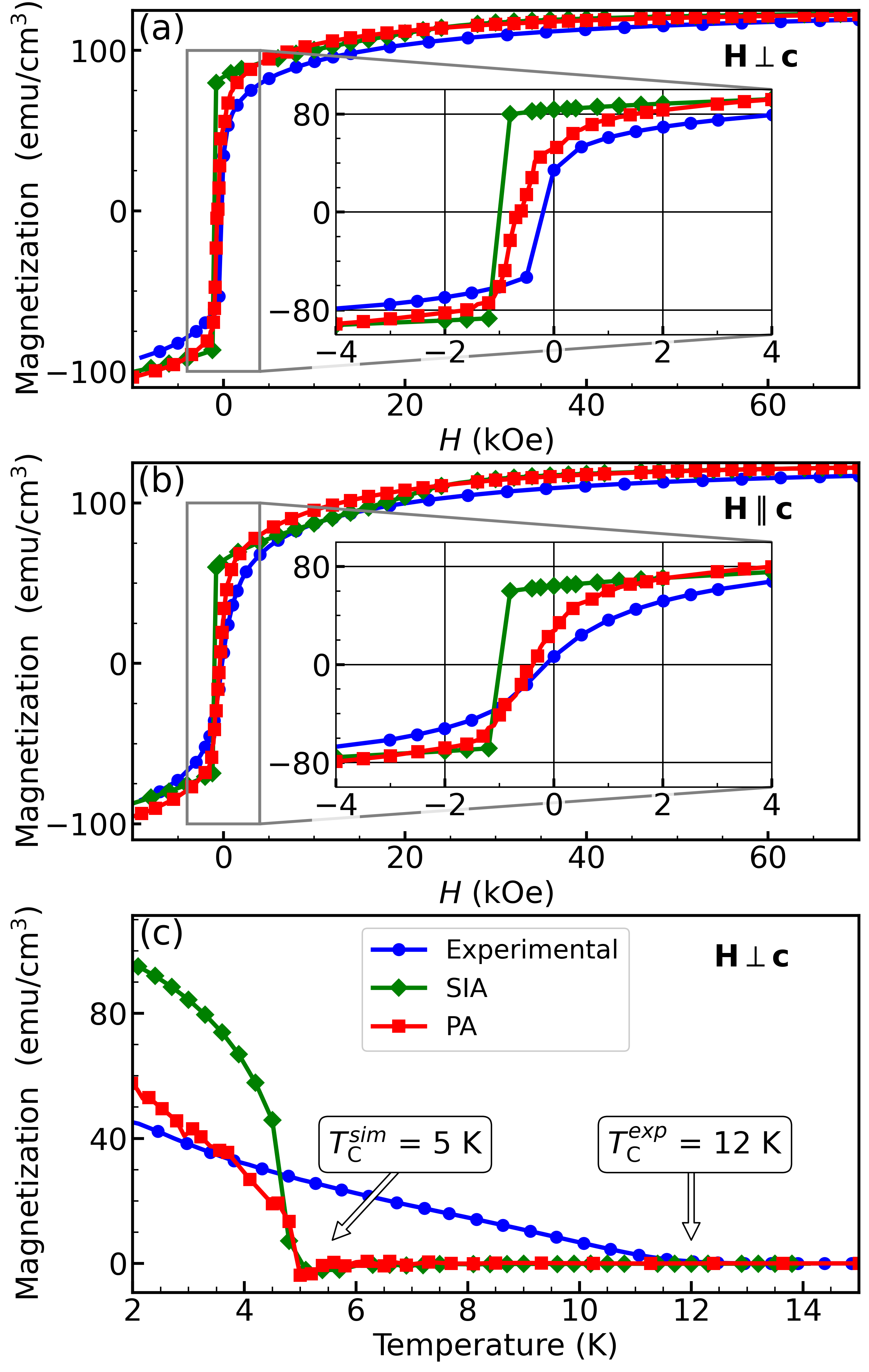}
	\caption{ Experimental and simulation results of magnetization and thermoremanent magnetization for the Ga$_{1-x}$Mn$_x$N sample with $x=7.9\%$. Magnetization measurements (blue dots) are performed at $T$ = 2 K for two magnetic field orientations: (\textbf{a}) in-plane ($\mathbf{H} \perp \mathbf{c}$) and (\textbf{b}) out-of-plane ($\mathbf{H} \parallel \mathbf{c}$). Insets show close-up on the weak magnetic field regime. For the clarity of the presentation only one branch of the hysteresis loop is shown. (\textbf{c}) The in-plane thermoremanent magnetization. The numerical atomistic spin model simulation results are obtained using single-ion anisotropy (SIA, green diamonds) and pair anisotropy (PA, red squares) approaches. }
	\label{fig:mag_curie}
\end{figure}
Having established that the magnetic anisotropy of nearest-neighbor Mn ion pairs deviates significantly from that of isolated Mn ions, this section examines whether a pair-based anisotropy model better aligns with experimental data for a representative ferromagnetic Ga$_{1-x}$Mn$_x$N sample with a typical concentration of $x=7.9\%$.
Two primary theoretical approaches exist to model the magnetic properties of Ga$_{1-x}$Mn$_x$N. In the extremely dilute limit, a quantum mechanical crystal field model (CFM) has proven effective in reproducing magnetization $M(H, T)$ curves as functions of magnetic field $H$ and temperature $T$~\cite{Herbich:1998, Stefanowicz:2010_PRB, Bonanni:2011_PRB, Devillers:2012_SR, Sztenkiel:2016_NatComm, Edathumkandy:2022_JMMM, Sztenkiel:2020_NJP, Sztenkiel:2022_JMMM}. This approach has successfully captured the magnetic~\cite{Bonanni:2011_PRB, Sztenkiel:2016_NatComm}, optical~\cite{Gosk2005}, and magnetoelectric~\cite{Sztenkiel:2016_NatComm} characteristics of Ga$_{1-x}$Mn$_x$N layers with Mn concentrations $x \leq 2.5\%$. 
The CFM framework is well-suited to describing isolated Mn ions and small magnetic clusters consisting of up to four Mn$^{3+}$ ions coupled by ferromagnetic superexchange~\cite{Edathumkandy:2022_JMMM, Sztenkiel:2020_NJP, Sztenkiel:2022_JMMM}. However, modeling larger systems necessitates classical approaches~\cite{Sztenkiel:2023_AdvTheorSim}. Recently, we applied the Landau--Lifshitz--Gilbert (LLG) formalism to simulate magnetization, magnetoelectric behavior~\cite{Sztenkiel:2025_CommMater}, and ferromagnetic resonance (FMR)~\cite{Edathumkandy:2025_PRB} in Ga$_{1-x}$Mn$_x$N with $x \approx 6\%$. This method represents spins as classical vectors and tracks their dynamic evolution. Despite general success, discrepancies between experimental and simulated $M(H)$ curves emerged at low end of cryogenic temperatures ($T \approx 2$~K). These deviations are attributed to the increasing likelihood of Mn clustering at higher concentrations, particularly the formation of NN Mn--Mn pairs. Therefore, to investigate the effect of pair anisotropy, we employ an atomistic spin model governed by the stochastic Landau--Lifshitz--Gilbert (sLLG) equation and/or the Monte Carlo (MC) approach.
A large simulation box with dimensions of $25 \times 25 \times 5~\mathrm{nm}^3$ is constructed, comprising $9\,860$ randomly distributed Mn spins substituting Ga cation sites in a wurtzite GaN lattice. The simulated Mn concentration $x = 7.9\%$ matches the experimental one. 
The investigated spin system is described by the following Hamiltonian: 
\begin{multline}
\label{Eq:Hamiltonian}
\mathcal{H} = 
\underbrace{- \sum_{\langle i,j \rangle} J_{ij} \mathbf{S}_i \cdot \mathbf{S}_j}_{\text{Exchange}} 
\overbrace{-\frac{1}{4} K^{TR} \sum_{i} \left[ S_{iz}^2 - \left( S_{ix}^2 + S_{iy}^2 \right) \right]}^{\text{Trigonal}}\\ + \mathcal{H}^{an}
\underbrace{- \mu_S \sum_i \mathbf{H} \cdot \mathbf{S}_i}_{\text{Zeeman}}
\end{multline}
where $\mathbf{S}_i$ is a normalized classical vector, corresponding to the local magnetic moment of $i$-th Mn ion. 
The Hamiltonian includes contributions from ferromagnetic superexchange interactions, magnetocrystalline anisotropy (trigonal term and $\mathcal{H}^{\mathrm{an}}$) and the Zeeman interaction (for details see appendix~\ref{app:atomistic}). To isolate and analyze the specific contribution of pair anisotropy, the anisotropy component of the Hamiltonian, $\mathcal{H}^{\mathrm{an}}$, is evaluated under two distinct scenarios:
\begin{itemize}
	\item \textbf{Single ion anisotropy (SIA):} Only Jahn--Teller anisotropy is included and applied uniformly to all Mn ion sites. The pair anisotropy is neglected. Then:
	
	\begin{equation}
	\mathcal{H}^{an} = -\frac{1}{2} K^{\mathrm{JT}} \sum_{i} \sum_{j=A,B,C} \left( \mathbf{S}_i \cdot \mathbf{e}^{\mathrm{JT}}_j \right)^4\\
	\end{equation}
	
	\item \textbf{Pair anisotropy (PA):} Both Jahn--Teller and pair anisotropies are included. However, the Jahn--Teller anisotropy is omitted for Mn ions forming nearest-neighbor pairs, based on the assumption that the presence of such a pair locally destroys JT effect. Then:
	
	\begin{multline}
	\label{Eq:PA_Hamiltonian}
	\mathcal{H}^{{an}} =\
	-\frac{1}{2} K^{\mathrm{JT}} \sum_{i \notin \text{pair}} \sum_{j=A,B,C} \left( \mathbf{S}_i \cdot \mathbf{e}^{\mathrm{JT}}_j \right)^4\\
	- K^{P} \sum_{i,j \in \text{pair}} \left( \mathbf{S}_i \cdot \mathbf{e}^{P}_{ij} \right)^2
	\end{multline}
	
	where $\mathbf{e}^{P}_{ij}$ is a vector connecting $i$-th and $j$-th spin. 
	
\end{itemize}

\new{For the sake of model simplicity in the atomistic simulations, we utilized the total calculated pair anisotropy without explicitly decomposing it into the symmetric anisotropic exchange and local environment-induced contributions. Since the total energy surface $\Delta E(\theta,\phi)$ used in the LLG code effectively accounts for both mechanisms, this approach maintains the physical consistency of the magnetization dynamics while avoiding excessive parametric complexity.}

\new{While our DFT investigations focus on Mn–Mn pairs due to the immense configurational complexity of larger clusters, our atomistic simulations explicitly account for the presence of triples and larger aggregates. This is achieved by decomposing the anisotropy of larger clusters into a sum of pairwise contributions (please see Eq.~\ref{Eq:PA_Hamiltonian}). For the experimental Mn concentration of $x$ = 0.079, where the probability of a Mn ion participating in a cluster larger than a pair reaches 23.9~$\%$, this decomposition ensures that the cumulative effect of multiple Mn neighbors is effectively captured within our model.}

Fig.~\ref{fig:mag_curie} compares the experimental data with simulation results. 
Below, we listed the best-fit parameters:
\begin{itemize}
	\item \textbf{SIA}: $J_{nn}$ = 4.0 meV, $K^{TR}$ = 0.05 meV,\\ $K^{\mathrm{JT}}$ = 0.75 meV.
	\item \textbf{PA}: $J_{nn}$ = 4.0 meV,$K^{TR}$ = 0.05 meV,\\ $K^{\mathrm{JT}}$ = 0.40 meV, $K^{P}$ = 0.75 meV.
\end{itemize}

In simulations we use the pair anisotropy model with a single $K^{P}$ and $K^{TR}$ parameters for all Mn--Mn pairs. This approach yields the best agreement with experiment. However, when we instead used the DFT-derived values (opposite signs of $K^{P}$ for in-plane vs out-of-plane pairs, as described in Sec.~\ref{Sec:MagneticAnisotropy}) the simulated magnetization deviated dramatically from the experimental data. This means that real material behaves as if the Mn--Mn pair anisotropy is smeared, possibly due to presence of higher order clusters or coupling beyond nearest neighbors. Thus, while the DFT fits reveal important local anisotropy nuances, the success of the uniform $K^{P}$ model indicates that effective modeling must balance local detail with realistic averaging of anisotropy contributions.
\new{It should be emphasized that the discrepancy between the $ab\ initio$ pair-anisotropy energy ($0.12$~meV) and the effective parameter used in our atomistic simulations ($0.75$~meV) is expected when transitioning from a $T=0$~K ground-state theory to a finite-temperature model. The simulation parameter serves as a renormalized effective constant that absorbs the influences of thermal fluctuations, potential lattice imperfections, the presence of larger clusters and long-range magnetic interactions, which are not captured in the idealized DFT supercell.}

As shown in Fig.~\ref{fig:mag_curie}, the PA model reproduces the experimental magnetization curves much more accurately, although the calculated critical temperature ($T_\mathrm{C}$) in both approaches remains approximately twice lower than the experimental value. 
The smoother shape of the $M(H)$ curves obtained within the PA model allows for the use of larger superexchange constants, which in turn increases the simulated $T_\mathrm{C}$.
In this study, however, we focus primarily on anisotropy effects, therefore, for comparison purposes the same $J_{ij}$ values are applied in both models.
Our goal is to reproduce the experimental $M(H, T)$ as accurately as possible, taking into account the overall shape of the curves, the saturation magnetization, and the coercive field values. 
According to Fig.~\ref{fig:mag_curie}, the approach based on PA provides a significantly better agreement with the experimental data. The SIA model exhibit noticeably overestimated values of remanent magnetization and coercive fields. Moreover, the overall shape of the $M(H)$ and $M(T)$ curves deviates significantly from the experimental data. These discrepancies are substantially reduced when PA model is applied: $M(H,T)$ curves more closely resemble the characteristic shape profile typically observed in Ga$_{1-x}$Mn$_x$N layers. 
%
%
\section{Summary}
%
%
We have identified and quantified a novel source of magnetic anisotropy in disordered systems, the \textit{pair-induced uniaxial anisotropy}, arising from the presence of nearest-neighbor magnetic ions. The effect is investigated in dilute magnetic semiconductor Ga$_{1-x}$Mn$_x$N. Our first-principles calculations reveal that such pairs break the local crystal-field symmetry and induce orientation-dependent magnetic energy contributions that are absent in isolated ions. This symmetry-breaking effect, captured by mapping the DFT-computed energy landscape onto a simplified spin Hamiltonian, highlights the need to move beyond the commonly used single-ion anisotropy approximation.

Importantly, our atomistic spin simulations demonstrate that inclusion of pair anisotropy leads to significantly better agreement with experimental magnetization curves, underscoring its critical role in accurately modeling the macroscopic magnetic behavior of disordered systems. Given the statistical likelihood of magnetic ion clustering in disordered systems, the pair anisotropy mechanism is expected to be relevant not only for Ga$_{1-x}$Mn$_x$N but also for a wider class of systems, including random magnetic alloys, cluster-based magnets, spin glasses, diluted magnetic oxides, and two-dimensional materials with randomly distributed spin centers.

These results advance the microscopic understanding of spin anisotropy in disordered magnetic systems and pave the way for more predictive multiscale modeling approaches linking first-principles calculations to mesoscopic magnetic behavior.

%
\section*{Data availability}
%
%
All data supporting the findings of this study  are available from corresponding author upon reasonable request. 
%
%
\section*{Acknowledgments}
%
%
The VASP output data were further processed and analyzed using VASPKIT\cite{Wang:2021_CPCOM}.
The work was supported by the National Science Centre (Poland) through projects OPUS (2018/31/B/ST3/03438) and PRELUDIUM Bis (2021/43/O/ST3/03280 - N.G.S.). This work was supported by the High Performance Computing (HPC) resources of Aix-Marseille University by the project Equip@Meso (ANR-10-EQPX-29-01). D.S., K.D. and R.H. thank the Erasmus program of the European Commission for supporting mutual visits.

%
\appendix

%
%
%
%
\section{Material and magnetometry}\label{app:experimental}
%
%
The numerical simulations are compared for wurtzite Ga$_{1-x}$Mn$_x$N sample with $x=7.9\%$, which was grown using plasma-assisted molecular beam epitaxy. The growth was performed along the c-axis on a thick, fully relaxed GaN buffer layer deposited on a sapphire substrate. \new{The thickness of Ga$_{1-x}$Mn$_x$N layer is 1~$\mu$m.} Previous extensive characterization confirmed the high structural quality of the samples and ruled out the presence of secondary phases or Mn-rich clusters~\cite{Sztenkiel:2016_NatComm, Gas:2018_JALCOM}. Magnetic measurements are carried out using a Quantum Design superconducting quantum interference device (SQUID) MPMS-XL magnetometer, following the protocol described in Ref.~\cite{Sztenkiel:2016_NatComm, Sawicki:2011_SST, Gas:2019_MST, Gas:2021_JAC, Gas:2022_Materials}. 
%
%
\section{Density functional theory}\label{app:dft}
%
%
Our first-principles calculations are carried out within the DFT framework, using the projector augmented-wave (PAW) method~\cite{Bloechl1994} as implemented in the Vienna Ab initio Simulation Package (VASP). A plane-wave energy cutoff of 520~eV is employed, along with a Gaussian smearing of 0.01~eV. The Brillouin zone is sampled using a $2 \times 2 \times 2$ Monkhorst--Pack $k$-point mesh~\cite{Monkhorst1976}. All calculations were spin-polarized and employed the generalized gradient approximation (GGA) in the Perdew--Burke--Ernzerhof (PBE) form for the exchange-correlation functional.
To obtain a detailed density of states (DOS), a denser $k$-point mesh of dimensions $11 \times 15 \times 7$,  centered at gamma point, is used.

The pseudopotentials are sourced from the standard VASP library. Given the particular challenges associated with Ga atoms in DFT simulations~\cite{Lilienfeld2008}, we treat the semicore $3d$ states of Ga as valence electrons to improve the accuracy and transferability of the results, albeit with increased computational expense. Geometry optimizations are performed until the total energy convergence criterion of $\Delta E = 10^{-7}$~eV is satisfied for both electronic state and ionic position. Then the geometry is further optimized until the force on each ion is reduced to less than 0.05 eV/\AA{}.

To construct simulation cells, we use $N_1 \times N_2 \times N_3$ expansions along the $\mathbf{a}$, $\mathbf{b}$ and $\mathbf{c}$ crystallographic axes of the conventional 4-atom wurtzite GaN unit cell. The initial $1 \times 1 \times 1$ GaN supercell is fully relaxed in both atomic coordinates and lattice parameters, yielding optimized constants of $|\mathbf{a}| = |\mathbf{b}| = 3.2174$~\AA{} and $|\mathbf{c}| = 5.2359$~\AA. This structure is then expanded to form a $3 \times 2 \times 2$ supercell containing 48 atoms. One Ga atom is substituted with a Mn ion to create a mono-doped $\mathrm{Ga_{1-x}Mn_xN}$ configuration, corresponding to $x \cong 2\%$.

As, during the epitaxial growth of Ga$_{1-x}$Mn$_x$N thin films~\cite{Gas:2018_JALCOM}, the in-plane lattice parameters $|\mathbf{a}|$ and $|\mathbf{b}|$ are constrained by the GaN buffer layer. Therefore, following our previous work~\cite{Sztenkiel:2016_NatComm}, the in-plane lattice constants are fixed to the theoretical value for GaN ($|\mathbf{a}| = |\mathbf{b}| = 3.2174$~\AA), while only the out-of-plane lattice parameter $|\mathbf{c}|$ and the internal atomic coordinates are allowed to relax.
A full spin-polarized structural relaxation is then performed on this system, maintaining the unit cell shape and resulting in $|\mathbf{c}| = 5.246$~\AA.

To study the effect of Mn--Mn interactions, we introduce a second Mn ion into a NN Ga site relative to the first Mn atom. The lattice parameters are held fixed to those of the relaxed mono-doped supercell to preserve a consistent doping concentration. Only the internal atomic coordinates are relaxed. Owing to the existence of two symmetry-inequivalent NN Ga positions, we consider both in-plane and out-of-plane Mn--Mn configurations.
%
%
\section{Atomistic spin model}\label{app:atomistic}
%
%
%
A large simulation box with dimensions of $25 \times 25 \times 5~\mathrm{nm}^3$ is constructed, comprising $124,820$ Ga cation sites in a wurtzite GaN lattice.
Among these, $9,860$ randomly distributed Mn spins substitute Ga ions, corresponding to a concentration of $x = 7.9\%$.
This specific concentration is chosen to allow for direct comparison with experimental data obtained from a sample of the same composition.
To mitigate finite-size effects, periodic boundary conditions are applied in the in-plane directions.

The spin Hamiltonian of Mn ions in GaN takes into account the Zeeman energy, Heisenberg spin-spin interactions between Mn spins, and the magnetocrystalline anisotropy energy: $\mathcal{H}=\mathcal{H}^{\mathrm{Exch}}+\mathcal{H}^{\mathrm{Trigonal}}+\mathcal{H}^{an}+\mathcal{H}^{\mathrm{Z}}$. The exchange interaction between Mn ions $\mathcal{H}^{\mathrm{Exch}}={- \sum_{\langle i,j \rangle} J_{ij} \mathbf{S}_i \cdot \mathbf{S}_j}$ is considered up to the 14th nearest neighbor, with the coupling strength parameterized as $J_{ij} = J_0 \exp(-R_{ij}/b)$, where $R_{ij}$ is the Mn--Mn distance and $b = 1.1~\mathrm{nm}$ is the characteristic decay length.
\new{The relatively large effective range of the interaction reflects the nature of superexchange coupling, which is mediated by $p$-$d$ hybridization with the nitrogen ligands and thus extends beyond the physical radius of the localized Mn $d$-orbitals.}
The trigonal anisotropy term $\mathcal{H}^{\mathrm{Trigonal}}=-\frac{1}{4} K^{TR} \sum_{i} \left[ S_{iz}^2 - \left( S_{ix}^2 + S_{iy}^2 \right) \right]$ favors an easy-plane configuration (i.e., a hard axis along the out-of-plane direction) for negative values of $K^{TR}$.
The remaining anisotropy components (Jahn-Teller and/or pair anisotropy) are included in $\mathcal{H}^{an}$.
The Jahn--Teller anisotropy is modeled as a fourth-order cubic anisotropy along three mutually orthogonal directions:
$\mathbf{e}_{A}^{\mathrm{JT}}=\left[ \sqrt{\frac{2}{3}}, 0, \sqrt{\frac{1}{3}}  \right]$, $\mathbf{e}_{B}^{\mathrm{JT}}=\left[-\sqrt{\frac{1}{6}}, -\sqrt{\frac{1}{2}} ,\sqrt{\frac{1}{3}}  \right]$, $\mathbf{e}_{C}^{\mathrm{JT}} = \left[-\sqrt{\frac{1}{6}}, \sqrt{\frac{1}{2}} ,	\sqrt{\frac{1}{3}}  \right]$.
The pair anisotropy is introduced as a uniaxial anisotropy aligned along the axis connecting each nearest-neighbor Mn--Mn pair.

To simulate the magnetization as a function of external magnetic field, particularly in the hysteresis regime (i.e., far from equilibrium), we employ the stochastic Landau--Lifshitz--Gilbert (sLLG) equation at the atomistic level:
\begin{equation}
\label{eq:LLG}
\frac{\partial \mathbf{S}_i}{\partial t} = - \frac{\gamma}{1 + \alpha_G^2}
\left[
\mathbf{S}_i \times \mathbf{H}_{i}^{eff} +
\alpha_G\, \mathbf{S}_i \times \left( \mathbf{S}_i \times \mathbf{H}_{i}^{eff} \right)
\right],
\end{equation}
where $\gamma$ is the gyromagnetic ratio and $\alpha_G$ is the Gilbert damping parameter. The magnetic moment of the Mn ion at site $i$ is given by $\mathbf{M}_i = \mu_S \mathbf{S}_i$, where $\mu_S = g \mu_B S$ is the saturation magnetic moment, with $g = 2$, $S = 2$, and $\mathbf{S}_i$ being a normalized classical vector. The effective magnetic field includes both deterministic and stochastic components:
\begin{equation}
\mathbf{H}_{i}^{eff} = -\frac{1}{\mu_S} \frac{\partial \mathcal{H}}{\partial \mathbf{S}_i} + \mathbf{H}_{i}^{th},
\end{equation}
where $\mathbf{H}_{i}^{th}$ is the thermal fluctuation field, given by:
\begin{equation}
\mathbf{H}_{i}^{th} = \mathbf{\Gamma}(t) \sqrt{\frac{2 \alpha_G k_B T}{\gamma \mu_S \Delta t}},
\end{equation}
with $\mathbf{\Gamma}(t)$ being a Gaussian random vector with zero mean, unit variance, and no temporal correlations. Equation~\eqref{eq:LLG} is numerically integrated using a Heun method and a fixed time step of $\Delta t = 5 \times 10^{-6}$~ns.

To compute $M(H)$ curves, the system is initially magnetized by applying a strong external field (70 kOe), which is then reduced in a stepwise manner. At each step, the sLLG equation is used to evolve the system to a steady-state magnetization, using the final configuration from the previous step as the initial state. This procedure is performed for both in-plane and out-of-plane field orientations.

In contrast, the Monte Carlo (MC) approach is used to investigate the thermodynamic equilibrium properties of the magnetic system. It is applied to calculate the zero-field magnetization as a function of temperature and to determine the critical temperature $T_\mathrm{C}$.

%
%
\section{Non-cubic Jahn-Teller axis}\label{app:non_cube}
%
%
\begin{figure}[h]
	\centering
	\includegraphics[width=0.96\linewidth]{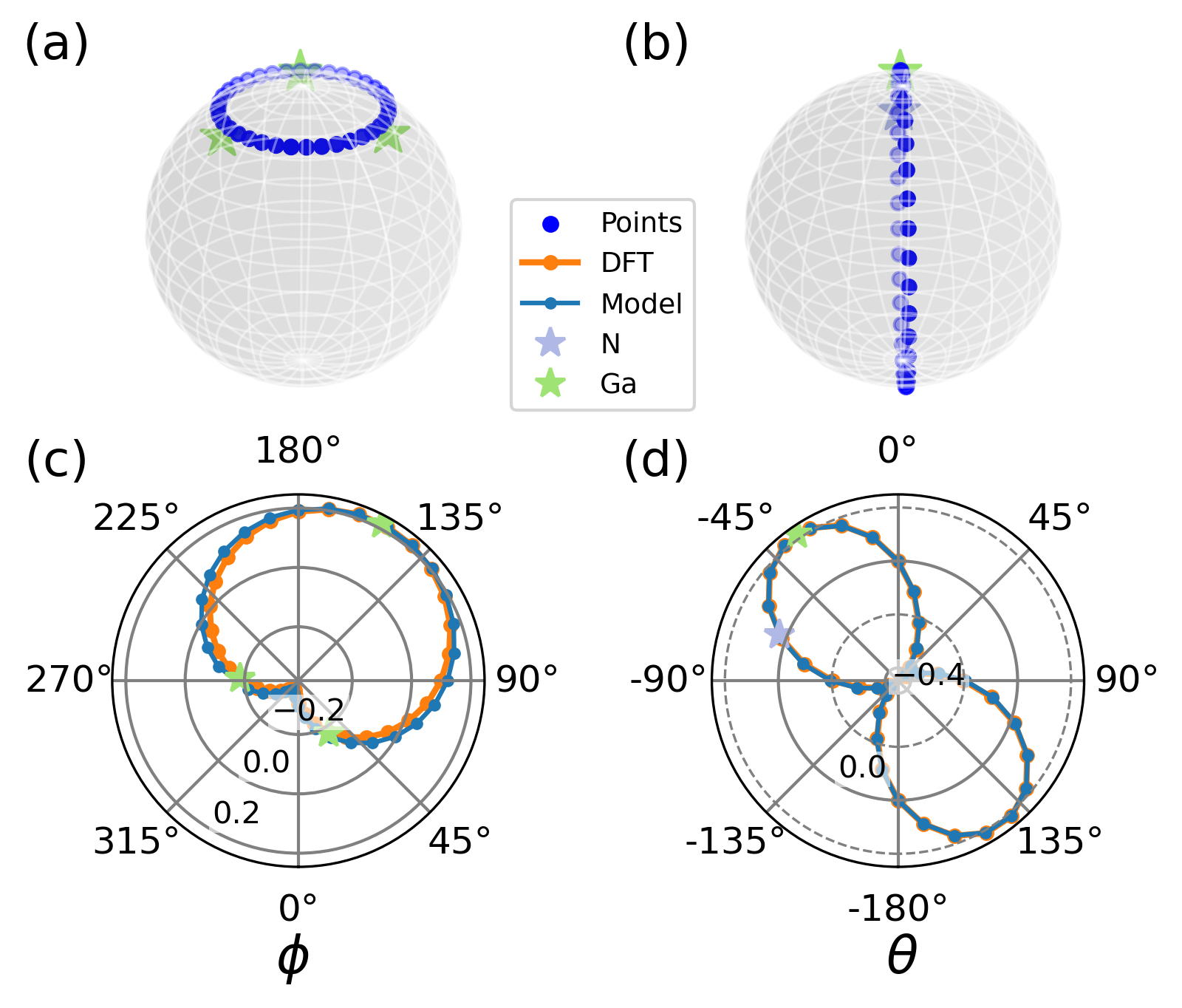}
	\caption{Magnetocrystalline energy of the supercell containing a single Mn ion. Panels (c) and (d) show the energy variation $\Delta E(\theta,\phi)$ obtained from DFT (orange lines) as a function of spherical angles $\phi$ and $\theta$, respectively. Blue lines represent fits to the spin Hamiltonian model defined in Eq. 2 in the main text. Spin orientations are illustrated in (a) and (b), corresponding to rotations over $\phi$ and $\theta$, respectively.}
	\label{fig:non_cube}
\end{figure}
Figure~\ref{fig:non_cube} presents the results of DFT simulations for the single-Mn supercell, along with a fit obtained using spin Hamiltonian assuming a non-cubic Jahn--Teller axis, $\mathbf{e}^{\mathrm{JT}}$, oriented at $\theta_{\mathrm{JT}}$ = $31.46^{\circ}$ and $\phi_{\mathrm{JT}}$ = $330^{\circ}$. The extracted anisotropy parameters are $K^{TR}$ = -1.00 meV/ion and $K^{\mathrm{JT}}$ = $0.67$ meV/ion. The angle $\theta_{\mathrm{JT}} = 31.46^{\circ}$, which yields the best fit, differs significantly from the cubic case, where $\theta_{\mathrm{JT}}$ = $54.73^{\circ}$.

%
%
\new{\section{Quenching of the Jahn-Teller effect in a Mn pair geometry.}\label{app:no_JT}}
%
%

\begin{table}[htb]
	\caption{\label{tab:in_pair_noJT} In-plane Mn-Mn pair geometry. Distances between nitrogen atoms surrounding the first and second Mn ions in the GaN supercell. N--N distance labels are provided in Fig.~\ref{fig:allsupercell}~\textbf{b}}
	\begin{ruledtabular}
		\begin{tabular}{ccccccc}
			N--N distance & $A$ & $B$ & $C$ & $\alpha$ & $\beta$ & $\gamma$ \\
			Unit & (\AA) & (\AA) & (\AA) & (\AA) & (\AA) & (\AA) \\
			\colrule
			first Mn  & 3.2141 & 3.2559 & 3.2334 & 3.2391 & 3.2221 & 3.247 \\
			second Mn    & 3.213 & 3.2166 & 3.2669 & 3.2369 & 3.2434 & 3.2113 \\
		\end{tabular}
	\end{ruledtabular}
\end{table}

\begin{table}[htb]
	\caption{\label{tab:out_pair_noJT} Out-of-plane Mn-Mn pair geometry. Distances between nitrogen atoms surrounding the first and second Mn ions in the GaN supercell. N--N distance labels are provided in Fig.~\ref{fig:allsupercell}~\textbf{b}}
	\begin{ruledtabular}
		\begin{tabular}{ccccccc}
			N--N distance & $A$ & $B$ & $C$ & $\alpha$ & $\beta$ & $\gamma$ \\
			Unit & (\AA) & (\AA) & (\AA) & (\AA) & (\AA) & (\AA) \\
			\colrule
			first Mn  & 3.2452 & 3.2585 & 3.251 & 3.2273 & 3.2294 & 3.2408 \\
			second Mn    & 3.238 & 3.2279 & 3.2361 & 3.2594 & 3.1757 & 3.21 \\
		\end{tabular}
	\end{ruledtabular}
\end{table}

\new{A detailed comparison of the N–N distances for isolated Mn ions (Tab.~\ref{tab:mono_mn_abc} in the main text) versus Mn–Mn pairs (shown in Tab.~\ref{tab:in_pair_noJT} and ~\ref{tab:out_pair_noJT}) reveals a fundamental change in the distortion symmetry. The isolated ions exhibit the typical tetragonal distortion pattern associated with the Jahn-Teller effect, characterized by two elongated and four shortened N–N distances as described in Eq.~\ref{eq:jt_para}. Specifically, Tab.~\ref{tab:mono_mn_abc} reveals that two elongated N-N distances are $B$ and $\beta$, whereas $A$ = $C \pm 0.07 \%$ and $\alpha$ = $\gamma \pm 0.14 \%$. 
On the contrary, the JT pattern is absent in the pair configurations. Analysis of the N-N distances shows that the local distortions for Mn--Mn pairs cannot be reconciled with a superposition of trigonal and tetragonal symmetries. This indicates that the local environment is shaped by the pair orientation rather than a standard Jahn-Teller effect, which would typically impose a well-defined tetragonal distortion. 
} 

%
%
\new{\section{Comparison with DFT+U}\label{app:Anizo_U}}
%
%

%
%
%
%
\begin{figure}[htb]
	\centering
	\includegraphics[width=0.96\linewidth]{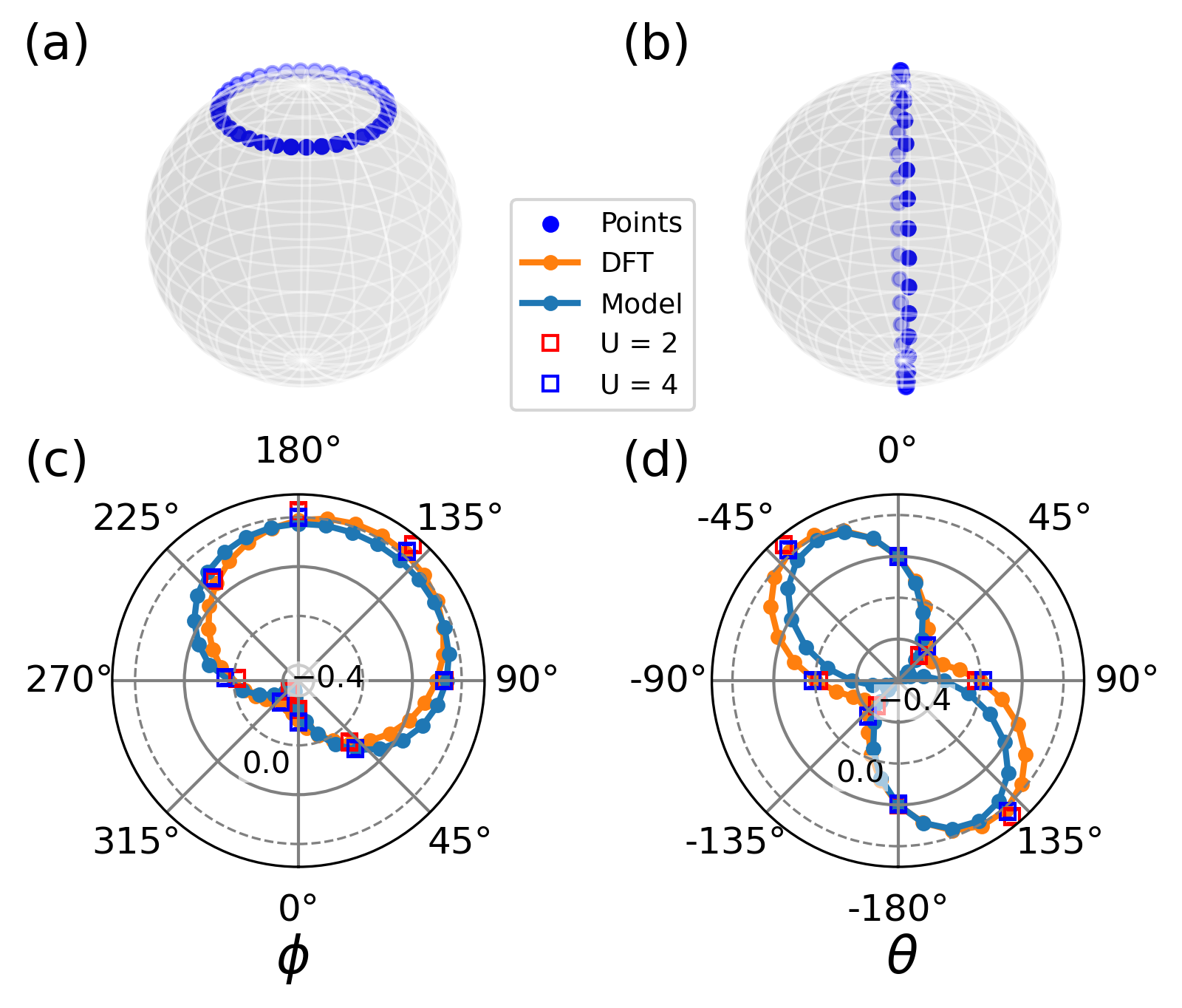}
	\caption{Magnetocrystalline energy of the supercell containing a single Mn ion. Panels (\textbf{c}) and (\textbf{d}) show the energy variation $\Delta E(\theta,\phi)$ obtained from DFT without $U$ correction (orange lines) as a function of spherical angles $\phi$ and $\theta$ of Mn spin, respectively. Blue lines represent fits to the spin Hamiltonian model defined in Eq.~\ref{eq:ani_model}. DFT results with $U=2$~eV and $U=4$~eV, calculated for selected spherical angles, are represented by red and blue squares, respectively.  Mn spin orientations are illustrated in (\textbf{a}) and (\textbf{b}), corresponding to rotations over $\phi$ and $\theta$, respectively.\label{fig:mono_fit_U}}
	
\end{figure}
\begin{figure*}[htb]
	\centering
	\includegraphics[width=\linewidth]{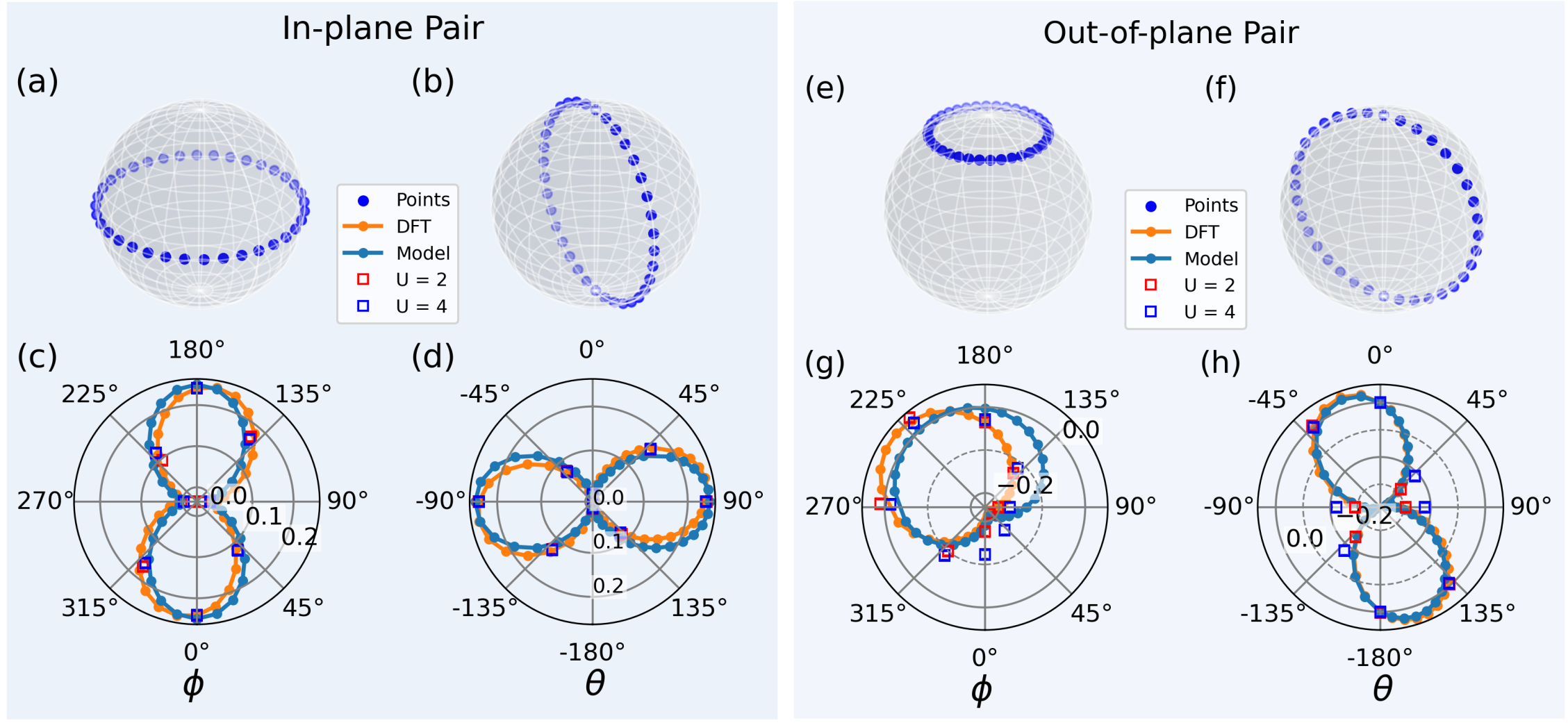}
	\caption{Magnetocrystalline energy of the supercell containing the in-plane (left panel) and out-of-plane (right panel) Mn--Mn pair. Results obtained from DFT without $U$ correction (orange) and spin Hamiltonian fit (blue) for both Mn spins rotations over $\phi$ and $\theta$ are shown in (\textbf{c}-\textbf{d}) and (\textbf{g}-\textbf{h}). DFT results with $U=2$~eV and $U=4$~eV, calculated for selected spherical angles, are represented by red and blue squares, respectively. The spin Hamiltonian model is defined in Eq.~\ref{eq:ani_model_pairs}. Panels (\textbf{a}), (\textbf{b}), (\textbf{e}) and (\textbf{f}) illustrate the corresponding Mn spins rotation geometries. \label{fig:pair_fit_U}}
\end{figure*}

\new{To evaluate the influence of strong electron correlation on the magnetic anisotropy energy (energy variation $\Delta E(\theta,\phi)$ as a function of the spherical angles $\phi$ and $\theta$ of the Mn spins), we performed benchmark calculations using the DFT+$U$ method. We applied the $U$ parameter to the Mn 3$d$ states with values of $U=2$~eV and $U=4$~eV. These calculations were carried out using the atomic positions previously relaxed with the PBE functional ($U=0$), allowing us to isolate the purely electronic effect of $U$ on the magnetic energy surface.
In Fig.~\ref{fig:mono_fit_U}, we present a comparison between previous PBE and new DFT+$U$ results for a single Mn ion. The results show that the functional form and the magnitude of the single-ion  anisotropy remain remarkably stable across the investigated range of $U$. Figure.~\ref{fig:pair_fit_U} shows the corresponding results for Mn--Mn pairs in both in-plane and out-of-plane configurations. We observe that the overall energy profiles $\Delta E(\theta,\phi)$ for $U=2$ and 4~eV are in very good agreement with the $U=0$ results. Although small numerical discrepancies are present, the positions of the energy minima and the relative barriers between different orientations are well-preserved. These findings confirm that the use of the PBE functional provides a physically accurate description of the magnetic anisotropy in Ga$_{1-x}$Mn$_x$N. The inclusion of $U$ does not significantly alter the pair-induced effects, justifying the numerical stability and reliability of the results presented in the main text}

%
%
%
%

%

\end{document}